# Changes in Household Net Financial Assets After the Great Recession: Did Financial Planners Make a Difference?

Joseph Goetz, Ph.D.

University of Georgia

Lance Palmer, Ph.D.

University of Georgia

Lini Zhang, Ph.D.

The Ohio State University

Swarn Chatterjee, Ph.D.

University of Georgia

# Changes in Household Net Financial Assets After the Great Recession: Did Financial Planners Make a Difference?

## Abstract

This study utilized the 2007-2009 Survey of Consumer Finances (SCF) Panel Data Set to examine the impact of financial planner use on household net financial asset level during the Great Recession. Data included 3,862 respondents who completed the SCF survey and a follow-up panel interview. The results indicated that starting to use a financial planner during the Great Recession had a positive impact on preserving and increasing the value of households' net financial assets, while curtailing the use of a financial planner during this time had a negative impact on preserving the value of households' net financial assets. Thus, study findings indicated that the benefit of using a financial planner may be particularly high during a major a financial downturn.

## Introduction

The early-21st century Great Recession wreaked havoc on many investors' financial portfolios. Based on longitudinal data of the Panel Study of Income Dynamics (PSID), Pfeffer, Danziger, and Schoeni (2013) reported that declines in net worth from 2007 to 2009 were large, and the declines continued through 2011. From 2007 to 2011, 12.2 percent of households experienced a loss of $250,000 or more in net worth, and 33.2 percent of households lost at least $50,000. According to Survey of Consumer Finances (SCF) 2007 to 2009 longitudinal data, high-wealth households were more exposed to shocks in financial markets through their ownership of retirement savings and other financial assets (Grinstein-Weiss, Key, & Carrillo, 2015). Throughout this economic crisis, many investors sought advice from professional financial advisors, and this advice likely played an important role in improving the financial wellness of those individuals (Grable & Joo, 1999).

Financial advice can be divided into five areas: (a) debt advice, (b) saving or investment advice, (c) mortgage or loan advice, (d) insurance advice, and (e) tax planning advice (FINRA, 2009). Among these areas, saving and investment is one of the most consequential yet daunting decisions the consumers face (Goldstein, Johnson, & Sharpe, 2008). Consistent with the 1994 Certified Financial Planners (CFP) Survey of Trends in Financial Planners and the 1994 International Association for Financial Planners (IAFP) Survey of Financial Advisors, Bae and Sandager (1997) found that consumers reported the use of a financial planner primarily for advice on retirement funding, investment growth, and tax planning.

Among the 2009 FINRA survey respondents, 56.7 percent reported receiving some form of financial advice, 8 percent reported obtaining advice on debt management, more than 33 percent reported obtaining advice on saving and investing, 23.5 percent reported receiving advice about a loan, about 33 percent reported obtaining advice on insurance, and about 21 percent reported working with a tax planning advisor (Collins, 2012). Based on 1998 SCF data, Elmerick, Montalto, and Fox (2002) found that 2.7 percent of households obtained advice from financial planners on only credit or borrowing, 11.5 percent looked for recommendations on only saving or investing, and 7 percent obtain both credit or borrowing and saving or investing. As households most often seek financial advice on saving and investing, this study focuses on the use of financial planners for saving and investment advice.

## Literature Review

*Use of Financial Planners*

Increasingly, people seek professional financial advice before making financial decisions. Elmerick, Montalto, and Fox (2002) reported that 21 percent of households used a financial planner and that the tendency to use financial planners increased with net worth and level of financial assets. Also using 1998 to 2007 SCF data, Hanna (2011) found that the use of a financial planner increased from 21 percent in 1998 to 25 percent in 2007, an estimated increase of almost five million households from just 2004 to 2007.

These financial planners tend to serve wealthy people. Finke, Huston, and Winchester (2011) found that the wealthy were more likely to pay for professional financial advice as well as comprehensive financial planning services when making financial decisions. Collins (2012) also found that individuals with high incomes, high educational attainment, and high levels of financial literacy were more likely to receive financial advice. The results of Hackethal, Haliassos, and Jappelli's (2012) study also revealed a strong positive correlation between using financial advice and wealth level.

While most previous financial advisor research focused on client and non-client differentiation using cross-sectional data, Cummings and James (2014a) were the first to focus their attention on the dynamic use of financial advisors by employing longitudinal data. Based on the 1993 and 1995 waves of Asset and Health Dynamics among the Oldest Old (AHEAD), the authors separated financial advisor use into four groups: no financial advisor, get financial advisor, drop financial advisor, and keep financial advisor. In this study, we also adopted the dynamic use of financial planners and divided them into four groups: (1) continue to use a financial planner, (2) start to use a financial planner, (3) discontinue to use a financial planner, and (4) never use a financial planner.

*Impact of Financial Planners*

Despite of the growing use of financial planners, relatively little is known about the impact that financial planners have on wealth. Much previous literature focused on studying *financial advisors* rather than *financial planners*, which often includes a broader group of professionals (e.g., broker, bankers, and insurance agents). In fact, the U. S. Financial Industry Regulatory Authority (FINRA) describes the main groups of investment professionals who may use the term *financial advisor* as the following: brokers, investment advisors, accountants, lawyers, insurance agents, and financial planners. Using data from the Asset and Health Dynamics Among the Oldest Old (AHEAD), Cummings and James (2014b) found that having a financial advisor was positively associated with subsequent net worth and investment returns. More specifically, their results showed that the positive impact of a financial advisor on subsequent wealth was likely due to a greater allocation to equities. Grable and Chatterjee (2014) used zeta to measure the value of advice in reducing wealth volatility, and the results showed that respondents who had previously met with a financial advisor experienced less wealth volatility on a risk-adjusted basis. However, little is known, regarding the impact of *financial planners* as a specific and distinct professional group. In contrast to previous research that focused on financial advisors, only financial planners were included in this study.

*Conceptual Framework*

Using 2007 and 2009 Survey of Consumer Finances (SCF) panel data, this research explores the impact of financial planner use on households' net financial assets during the Great

Recession. This study contributes to the literature in several ways. First, different from previous research that focused on financial advisors, this study focuses on the association between financial planner use and households' net financial assets. Second, this study examines whether the use of financial planners made a difference in preserving and increasing wealth during the Great Recession, the period that many households were exposed to high losses in financial assets. Third, instead of studying the use of financial planners on a cross-sectional basis, this study explores use of financial planners based on longitudinal data. Finally, this study focuses on high-wealth households, which were both more likely to use financial planners and to be exposed to substantial financial shocks during recession.

The study uses a personal finance help-seeking behavior model as its conceptual framework. Suchman (1966) was among the first theorists to develop a conceptualization of help-seeking behaviors to explain socio-psychological behaviors. Based on his work, Grable and Joo (1999) developed the personal finance model as the framework to explain and predict financial help-seeking behavior.

Help-seeking behavior has been defined as an action by an individual or a household to seek assistance from a secondary source (Grable & Joo, 1999). In this study, help-seekers search for a financial planner to help them preserve and increase the value of their net financial assets. The process of the framework is straightforward and consists of five steps:

Step 1 – the exhibition of a personal financial behavior(s)

Step 2 – the evaluation of the financial behavior(s)

Step 3 – the identification of financial behavior(al) causes

Step 4 – a decision to seek help

Step 5 – a choice between help provider alternatives

And finally, evaluating the outcomes of the help-seeking behavior completes the process. A person's demographic and socioeconomic characteristics influence the outcomes of financial help-seeking behavior (Joo & Grable, 2001; Salter et al., 2010), therefore these characteristics will be controlled when examining the outcomes of financial planner use in this study.

Under the personal finance help-seeking behavior framework, during Step 1 of the process, an individual or a household observed the substantial loss of value in stocks, mutual funds, and other investments and may have made the decision to sell investments to avoid greater loss. During Step 2, the individual or household began to calculate and evaluate their market loss and tried to find the cause for the loss. At Step 3, some individuals or households identified the loss in their financial assets as the result of their panic and irrational selling behavior. In Step 4, some individuals or households considered seeking financial help from professionals. Step 5 shows that some people or households finally chose to seek out the assistance of a professional financial planner, while others might decide to seek help from friends, family members, colleagues, and other sources.

The final process is to evaluate to what extent seeking financial planner assistance helped individuals or households preserve or increase their net financial assets during and after the Great Recession. In this study, we focus on the influence of Step 5 of the personal finance help-seeking framework.

## Methodology

### *Data and Sample*

This study uses the 2007-2009 Survey of Consumer Finances (SCF) Panel Data Set for data analysis. The SCF is a triennial survey of U.S. families, which includes information on families' balance sheets, pensions, income, and demographic characteristics. The SCF is sponsored by the Board of Governors of the Federal Reserve System in cooperation with the Statistics of Income Division of the Internal Revenue Service. The panel data set is based on the 2007 SCF and an interview with eligible respondents in 2009. The interview was motivated by a desire to understand more deeply the effects of the financial crisis on U.S. households. SCF conducted interviews for the 2009 survey between July and December of 2009, but a small number were finished in January 2010. National Opinion Research Center (NORC), a social science research center at the University of Chicago, collected the 2007-2009 SCF panel data.

The SCF unit of analysis is the household. The great majority of the survey focuses on the primary economic unit, which includes all people in the household who were economically interdependent with the respondent and/or his or her spouse or partner. For the 2007 survey, the respondent is the economically dominant single individual or the financially more knowledgeable member of the couple. There were 4,422 households that participated in the 2007 SCF survey and 3,862 of them completed a follow-up panel interview in 2009.

Missing data were a substantial problem in the SCF. A multiple imputation procedure used in the SCF yielded five values for each missing value, and the imputations were stored as five successive replicates of each data point recorded. Thus, the number of observations in the full data set (19,310) were five times the actual number of respondents (3,862). Only the first replicate of data for each observation was used for this study. Because the SCF over-sampled relatively wealthy households, use of appropriate weights was important for obtaining unbiased population estimates from the data.

Only high-wealth households were included as the sample for this study because they were most likely to use financial planners and to be more exposed to financial shocks during recession as mentioned earlier. Thus, the sample only includes those who fell into the fifth quintile of net financial assets. The final sample of 766 households from 2007-2009 SCF panel data were used in this study.

*Variables and Data Analysis*

This study uses multiple regressions to examine whether households that utilized financial planners preserved and increased the value of their net financial assets over households who did not seek the services of a financial planner. The dependent variable in this study is the percentage change of households' net financial assets from 2007 to 2009. The most important independent variable is use of financial planners in both 2007 and 2009. As mentioned earlier, use of financial planners includes four groups: (1) continue to use a financial planner, (2) start to use a financial planner, (3) discontinue to use a financial planner, and (4) never use a financial planner.

This analysis controls for several variables expected to influence change in net financial assets, including age, gender, marital status, race, education, household income, and household size. Quick emergency funds, including checking, savings, and money market accounts, also play a role in the change in net financial assets. Several behavior factors expected to influence change in net financial assets are captured by whether various loan or mortgage payments were made on time, whether spending exceeded households' income, risk tolerance level, change in homeownership, and smoking (as a proxy for self-regulation)

We use multiple regression to model the factors that may have an impact on households' net financial assets change. The following model is used for the data analyses:

$$y = \beta_0 + \beta_1 x_1 + \beta_2 x_2 + \beta_3 x_3 + \varepsilon$$

where y is the percentage change in households' net financial assets from 2007 to 2009. Each of the independent variables represents those factors that may be related to change in net financial assets. In the model, $x_1$ represents financial planner use. Financial planner use is divided into four groups: (1) those who had a financial planner in both 2007 and 2009, (2) those who had a financial planner in 2007 but dropped the financial planner in 2009, (3) those who did not have a financial planner in 2007 but started to use a financial planner in 2009, and (4) those who did not have a financial planner in either 2007 or 2009. In the regression model, the reference group is those who did not have a financial planner in both periods. To check the influence of behavioral variables that may impact the percentage change of net financial assets, $x_2$ includes variables such as saving behavior (measured by quick emergency fund ratio), spending behavior (whether spending exceeded households' income), payments scheduled (whether various loan or mortgage payments were made on time), risk tolerance, change in homeownership, and smoking (as a proxy for self-regulation). Finally, $x_3$ in the model is a vector of demographic variables that may influence the percentage change in net financial assets. Age, gender, marital status, race, education, household income, number of children are examined in the regression model. The behavioral variables and demographic variables are based on the responses in 2007. The error term is assumed to be normally distributed with a mean of zero.

## Results

### *Descriptive Statistics*

Table 1 reports both unweighted and weighted statistics of financial planner use before and after the Great Recession. The frequency and unweighted percentage show the use of financial planners for the sample selected from SCF data set. Because the SCF data set oversampled relatively wealthy households, suggested weights have been used to make the sample representative of the U.S. population.

Based on the weighted statistics provided in Table 1, around 31 percent of households had a financial planner in both 2007 and 2009, more than 16 percent of households had a financial planner in 2007 but dropped their financial planners in 2009, about 17 percent of household did not have a financial planner in 2007 and started to use a financial planner in 2009, more than 35 percent of the households did not have a financial planner in either 2007 or 2009.

Table 1

*Use of financial planners*

| Use of financial planners | Frequency | Unweighted Percent | Weighted Percent |
|---|---|---|---|
| Yes 07 Yes 09 | 248 | 32.38 | 30.95 |
| Yes 07 No 09 | 140 | 18.28 | 16.67 |
| No 07 Yes 09 | 124 | 16.19 | 16.97 |
| No 07 No 09 | 254 | 33.16 | 35.41 |

*Note.* N=766 for frequency and unweighted percent. Yes 07 Yes 09 represents those who had a financial planner in both 2007 and 2009, Yes 07 No 09 represents those who had a financial planner in 2007 but dropped use in 2009, No 07 Yes 09 represents those who did not have a financial planner in 2007 but started to use one in 2009, No 07 No 09 represents those who did not have a financial planner in both 2007 and 2009.

Table 2 presents the weighted descriptive statistics of additional variables. Percentage change in net financial assets, age, years of education, income, number of children, and quick emergency ratio are continuous variables, while race, gender, marital status, spending behavior, payment scheduled, smoking, risk tolerance, and change in homeownership are categorical variables. In Table 2, mean and median are reported for continuous variables, and percentages are reported for categorical variables. Except for the percentage change in net financial assets and change in homeownership between 2007 and 2009, all other variables are from cross-sectional data in 2007.

Financial assets in this study include transaction accounts or liquid assets: CDs, savings, money market deposit accounts, money market mutual funds, bonds, stocks, pooled investment funds or non-money market mutual funds, retirement accounts, cash value life insurance, other managed assets (including personal annuities and trusts with an equity interest and managed investment accounts), and other financial assets (a heterogeneous category including oil and gas leases, futures contracts, royalties, proceeds from lawsuits or estates in settlement, and loans made to others).

The quick emergency fund ratio is calculated by dividing quick emergency funds by monthly living expenses. Because the SCF data set provides income rather than expense data, most previous research has used quick emergency funds divided by monthly income to measure quick emergency fund ratio (Bhargava & Lown, 2006). This method is also applied to this study. Quick emergency funds include money in checking, savings, and money market accounts.

Table 2

*Descriptive Statistics in 2007(Weighted)*

| Variables | 2007 | |
|---|---|---|
| | Mean | Median |
| Net Financial Assets ($) | 4,561,097 | 2,754,690 |
| NFA Percentage Change | -22.85% | -27.00% |
| Age | 61.15 | 60 |
| Years of Education | 16.07 | 16 |
| Income ($) | 778,864.73 | 425,990.52 |
| Number of Children | 0.59 | 0 |
| Quick Emergency Fund Ratio | 8.57 | 2.88 |
| | Percent (%) | |
| Race | | |
|   White | 95.22 | |
|   Non-White | 4.78 | |
| Gender & Marital Status | | |

|   |   |
|---|---|
| Single Male | 8.91 |
| Single Female | 4.16 |
| Married | 86.94 |
| Spending Behavior |   |
| Spend More Than Income | 10.36 |
| Spend No More Than Income | 89.64 |
| Payment Scheduled |   |
| On Time | 97.62 |
| Not On Time | 2.38 |
| Smoking |   |
| Yes | 5.09 |
| No | 94.91 |
| Risk Tolerance |   |
| Risk Averse | 6.80 |
| Willing to Take Risk | 93.20 |
| Homeownership |   |
| Always Have House | 96.83 |
| Always Do Not Have House | 1.04 |
| Sell House | 1.53 |
| Buy House | 0.60 |

*Note.* N = 766. NFA Percentage Change represents Net Financial Assets Percentage Change.

Except for the descriptive statistics of the whole sample, descriptive statistics are presented in four groups by use of financial planner between 2007 and 2009 (see Table 3). From the descriptive statistics in 2007, we can see the differences of percentage change in net financial assets between four groups. The mean and median of percentage change in net financial assets are the smallest for those who did not have a financial planner in 2007 and started to use one in 2009, then followed by the group of those who had a financial planner in both 2007 and 2009.

Table 3

*Descriptive Statistics in 2007 in Four Groups (Weighted)*

| Variables | Yes 07 Yes 09 | Yes 07 No 09 | No 07 Yes 09 | No 07 No 09 |
|---|---|---|---|---|
| | Mean (Median) | Mean (Median) | Mean (Median) | Mean (Median) |
| Net Financial Assets ($) | 5,019,687 (2,656,308) | 4,973,757 (3,556,139) | 3,739,801 (2,635,596) | 4,359,405 (2,647,213) |
| NFA Percentage Change | -21.45% (-27.00%) | -30.97% (-30.76%) | -5.75% (-17.27%) | -28.45% (-33.46%) |
| Age | 59.21 (60) | 59.03 (60) | 62.91 (61) | 63 (61) |
| Years of Education | 16.31 (17) | 15.99 (16) | 16.33 (16) | 15.77 (16) |
| Income ($) | 857,105.50 (479,239.34) | 1,007,550.01 (490,954.08) | 728,378.66 (579,347.11) | 626,970.73 (240,684.64) |
| Number of Children | 0.76 (0) | 0.65 (0) | 0.48 (0) | 0.48 (0) |
| Quick Emergency Fund Ratio | 4.80 (2.34) | 7.63 (3.00) | 5.69 (1.94) | 13.69 (4.45) |
| | Percent (%) | Percent (%) | Percent (%) | Percent (%) |
| Race | | | | |
|   White | 92.59 | 92.94 | 99.70 | 96.45 |
|   Non-White | 7.41 | 7.06 | 0.30 | 3.55 |
| Gender & Marital Status | | | | |
|   Single Male | 10.56 | 6.47 | 5.07 | 10.42 |
|   Single Female | 2.04 | 5.91 | 2.95 | 5.78 |
|   Married | 87.40 | 87.62 | 91.98 | 83.80 |
| Spending Behavior | | | | |
|   Spend More Than Income | 10.42 | 14.64 | 7.74 | 9.53 |
|   Spend No More Than Income | 89.58 | 85.36 | 92.26 | 90.47 |
| Payment Scheduled | | | | |
|   On Time | 97.52 | 99.06 | 97.78 | 96.96 |
|   Not On Time | 2.48 | 0.94 | 2.22 | 3.04 |
| Smoking | | | | |
| Yes | 5.06 | 3.69 | 4.44 | 6.08 |
| No | 94.94 | 96.31 | 95.56 | 93.92 |
| Risk Tolerance | | | | |
|   Risk Averse | 0.89 | 17.55 | 5.33 | 7.62 |
|   Willing To Take Risk | 99.11 | 82.45 | 94.67 | 92.38 |
| Homeownership | | | | |
|   Always Have House | 98.60 | 97.50 | 96.50 | 95.13 |
|   Always Do Not Have House | 1.33 | 0 | 0.20 | 1.68 |
|   Sell House | 0 | 0 | 3.30 | 2.73 |
|   Buy House | 0.06 | 2.50 | 0 | 0.46 |

*Note. N* = 766. Yes 07 Yes 09 represents those who had a financial planner in both 2007 and 2009, Yes 07 No 09 represents those who had a financial planner in 2007 but dropped use in 2009, No 07 Yes 09 represents those who did not have a financial planner in 2007 but started to use one in 2009, No 07 No 09 represents those who did not have a financial planner in both 2007 and 2009. NFA Percentage Change represents Net Financial Assets Percentage Change.

*Multiple Regression*

A multiple regression model is used to explore the impact of financial planner use on the percentage change in net financial assets. The dependent variable is percentage change in net financial assets during the Great Recession, the independent variables include use of financial planners between 2007 and 2009, and control variables include a number of socioeconomic, demographic, and behavioral factors.

Table 4 shows the result from multiple regression. Compared to those who did not have a financial planner in both 2007 and 2009, the net financial assets for those who did not have a financial planner in 2007 but started to use one in 2009 had increased by 12.2 percent; the net financial assets for those who had a financial planner in 2007 but dropped to use in 2009 had decreased by 13.5 percent. There were no significant differences between those who had a financial planner in both 2007 and 2009 and those who did not have a financial planner in both periods. Compared to those who spent no more than available income, the net financial assets of those who spent more than income had decreased by 16.5 percent.

Table 4

*Multiple Regression Model on Percentage Change in Net Financial Assets*

| Variable | Coefficients | Std. Error | t | Sig. |
| --- | --- | --- | --- | --- |
| (Constant) | -2.462*** | 0.403 | -6.11 | 0.000 |
| Yes 07 Yes 09 | -0.051 | 0.050 | -1.02 | 0.308 |
| No 07 Yes 09 | 0.122* | 0.058 | 2.09 | 0.037 |
| Yes 07 No 09 | -0.135* | 0.060 | -2.27 | 0.024 |
| Quick Emergency Fund Ratio | -0.002 | 0.001 | -1.85 | 0.064 |
| Payment On Time | 0.149 | 0.136 | 1.10 | 0.272 |
| Smoking | -0.059 | 0.091 | -0.65 | 0.514 |
| Spend More Than Income | -0.165* | 0.066 | -2.49 | 0.013 |
| Risk Averse | 0.059 | 0.086 | 0.69 | 0.494 |
| Income (Log) | 0.155*** | 0.022 | 7.17 | 0.000 |
| Age | -0.006* | 0.002 | -2.44 | 0.015 |
| Years of Education | 0.035* | 0.015 | 2.36 | 0.018 |
| White | -0.009 | 0.095 | -0.10 | 0.923 |
| Single Female | -0.043 | 0.126 | -0.34 | 0.733 |
| Married | -0.012 | 0.070 | -0.17 | 0.863 |
| Number of Kids | -0.089*** | 0.024 | -3.70 | 0.000 |
| Always Do Not Have House | -0.006 | 0.194 | -0.03 | 0.977 |
| Sell House | 0.372* | 0.159 | 2.34 | 0.017 |
| Buy House | -0.274 | 0.255 | -1.07 | 0.285 |

*Note.* $N = 766$. Yes 07 Yes 09 represents those who had a financial planner in both 2007 and 2009, Yes 07 No 09 represents those who had a financial planner in 2007 but dropped to use in 2009, No 07 Yes 09 represents those who did not have a financial planner in 2007 but started to use one in 2009, No 09 No 09 represents those who did not have a financial planner in both 2007 and 2009.
*$R^2 = 17.06\%$*
*$p < .05$. ***$p < .001$.

The results also show that a 100 percent increase in income significantly increased the net financial assets by 15.5 percent. A one-year increase in age significantly decreased the net financial assets by 0.6 percent, while a one-year increase in years of education significantly increased the net financial assets by 3.5 percent. One more child in a household significantly decreased the household's net financial assets by 8.9 percent. Compared to those who had a house in both periods, those who sold their house during the recession increased the household's net financial assets by 37.2 percent.

## Conclusion

The results of this study indicate that starting to use financial planners during the Great Recession had a positive impact on preserving and increasing the value of households' net financial assets, while dropping financial planners had a negative impact on preserving the value of households' net financial assets. As such, the benefit of using a financial planner may be particularly high during a major a financial downturn.

Although no significant relationship was found between continued use of a financial planner and change in households' net financial assets, that does not mean keeping financial planners had no impact on preserving and increasing the value of net financial assets. For example, households who had a financial planner before the recession may have already been in a well-diversified portfolio and positioned to optimally rebalance and recover from a major drop in the stock markets; thus, the Great Recession did not have a significant impact on their net financial assets. In other words, keeping financial planners helped to avoid a potential loss in household net financial assets over this time period. It is also worth noting that the relationship between the use of financial planners and households' net financial assets is not a causal relationship. Thus, it is possible that households who experienced a large drop in their financial assets may have decided to fire their own financial planners after the Great Recession.

Multiple factors may be have contributed positive outcomes associated with hiring and retaining a financial planner during a recessionary period. For example, financial planners may provide a buffer to the well-documented emotional and cognitive biases that consumers experience, particularly during time of major stock market movements, that often to lead to sub-optimal investment decisions, and in turn, decreased wealth (Chatterjee & Goetz, 2017; Goetz & Gale, 2014; Goetz & James, 2008). It could also be that financial planners were more effectively rebalancing portfolios as the stock market plummeted (i.e., moving from bond to stock positions) than those consumers who dropped their financial planner, thus capturing greater wealth as the market began to recover in 2009. In summary, the use of financial planners was associated with preserving and increasing households' net financial assets during the Great Recession. Based on these findings, households should generally be encouraged to use financial planners and retain their financial planners during recessionary periods.